\documentstyle[12pt,epsfig]{article}

\newcommand{\be}{\begin{equation}}
\newcommand{\ee}{\end{equation}}

 1
\font\elevenrm=cmr10 scaled\magstep 1
 1

\def\reff{\hang\noindent}
\def\msol{{M}_{\odot}}

\def\lesssim{\,\lower 1mm \hbox{\ueber{\sim}{<}}\,}
\def\grsim{\,\lower 1mm \hbox{\ueber{\sim}{>}}\,}
\def\ueber#1#2{{\setbox0=\hbox{$#1$}%
  \setbox1=\hbox to\wd0{\hss$ #2$\hss}%
  \offinterlineskip
  \vbox{\box1\box0}}{}}
\def\ltsima{$\; \buildrel < \over \sim \;$}
\def\simlt{\lower.5ex\hbox{\ltsima}}

\begin{document}
\vspace*{1.8cm}
  \centerline{\bf CLUSTERS OF GALAXIES: DIAGNOSTIC TOOLS FOR COSMOLOGY}
\vspace{1cm}
  \centerline{Sabine Schindler}
\vspace{1.4cm}
  \centerline{Liverpool John Moores University}
  \centerline{\elevenrm Astrophysics Research Institute, Twelve Quays
House, Birkenhead CH41 1LD, U.K.}
\vspace{3cm}
\begin{abstract}
Clusters of galaxies can be used for very different kinds of
cosmological tests. I review a few of the methods: determination of
cluster masses and dark matter content, metal enrichment and its connection to
the origin of the intra-cluster gas, and the dynamical state of clusters.
The progress in the different fields expected from observations with
the new X-ray satellites XMM and Chandra is pointed out.
\end{abstract}
\vspace{2.0cm}

\section{ Introduction }

In recent years clusters of galaxies have been shown to be excellent
diagnostic tools for cosmological research. They provide a variety of
different ways to determine cosmological parameters like the mean density
$\Omega$ of the universe, 
the baryon density $\Omega_{baryon}$, the cosmological
constant $\Lambda$  or the Hubble constant $H_0$. 
Galaxy clusters are special for several reasons. (1) They are the
largest bound structures in the universe. With sizes of a few Mpc and
masses around $10^{15} \msol$ they constitute already a considerable
fraction of the universe. (2) They are
closed systems. As no matter can leave the cluster potential well the
mix of baryons and non-baryons should be
representative for the universe as a whole. 
(3) The crossing time -- the time it
takes a galaxy to move from one end of the cluster to the other -- is
of the order of a Hubble time. Therefore the information of the
formation process is not completely wiped out, but there are still
traces left, which provide insights into the early universe. (4)
Clusters are observable out to large redshifts. Comparison of
properties of distant and nearby clusters therefore yields 
information about
evolutionary effects, which are predicted to be different in different
cosmological scenarios. (5) They can be used as tracers
for the mass distribution in the universe, which is another quantity
to constrain cosmological models.

\section{Cluster components}

Clusters of galaxies consist of various components. Clusters were first
discovered as associations of hundreds to thousands of galaxies. These
galaxies are not at rest in the potential well, but move around with
velocities of the order of 1000 km/s. The galactic mix of morphological
types of galaxies in clusters differs from the mix in the field
(Dressler 1980): in 
clusters an excess of elliptical galaxies is visible, 
which is an indication of
interaction. This interaction can be between galaxies or between
galaxies and another component: the intra-cluster gas.

This gas fills all
the space between the galaxies. The density of the gas is relatively
low with $10^{-2} - 10^{-4}$ particles per cm$^{-3}$, but the temperature is
high at $10^7 - 10^8$ K. Such high temperatures are in good
agreement with the depth of the potential wells of clusters. Gas with these
specifications emits thermal bremsstrahlung in the X-ray range,
which makes it a particularly interesting cluster component because we can
observe it with X-ray satellites. These observations provide spatial
information about the morphology of the clusters as well as spectral
information which can be used to determine gas temperatures.
The X-ray spectra also show lines which
correspond to metallicities of about one third of the solar
value (Fukazawa et al. 1998). This is an indication that the gas
cannot be only of primordial 
origin, but at least part of it must have been processed in cluster
galaxies and expelled later from the galaxy potential wells.

Another interesting component is relativistic particles.
Their presence can be inferred from radio
observations of their synchrotron emission. 

These different components can be used to determine cosmological
parameters in different ways. In the following I present a selection
of particularly interesting methods.

\section{Cluster masses, baryon fraction and dark matter}

The mass fraction of the intra-cluster gas is not negligible. It makes
up about 10 - 30 \% of the total cluster mass. Compared to that the
mass in the galaxies is much smaller, about 3 - 5 \%. The major
fraction of the
mass is not visible and is therefore called dark matter. Hence 
measurements of the total mass of the clusters indirectly yield 
information on
the amount and the distribution of the dark matter (for an example see
the mass profile of the Virgo cluster in Fig.~1).

\begin{figure}[thb]
 \begin{center}
  \mbox{\epsfig{file=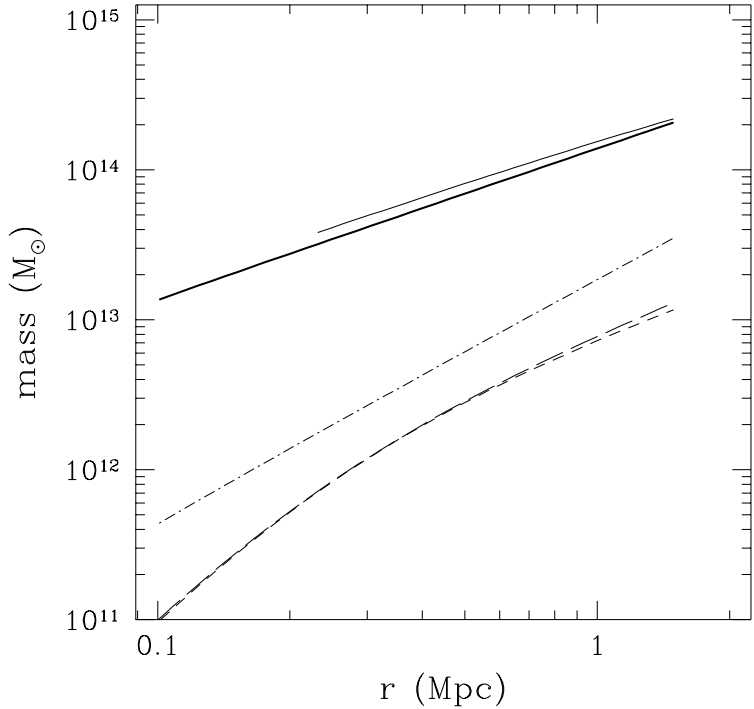,width=10cm}}
\vskip 0.7cm
  \caption{\em 
Integrated mass profile of the Virgo cluster around M87. Shown are the gas mass
  (dash-dotted line), the galaxy mass (calculated with two slightly
  different assumptions: short dashed line and long dashed line), as
  well as the total (gravitating) mass assuming  isothermal cluster gas
(bold solid line) and assuming a temperature gradient 
(thin solid line). For details see Schindler et al. (1999).
}
 \end{center}
\end{figure}

The total mass of clusters can be measured in different ways.
\begin{itemize}
\item The X-ray emitting gas can be used as a tracer of the cluster
potential. 
The depth and the shape of the potential well can be determined from
X-ray observations alone
via the temperature and density gradient of the
intra-cluster gas with the assumption of hydrostatic equilibrium.
\item A second method is based on the gravitational lensing
effect: light from galaxies behind clusters is deflected by the large
mass of the cluster, so that we see distorted images of these
galaxies. The giant arcs (strong lensing) as well as the statistical
distortion of all background objects (weak lensing) can be used for
the mass determination (Wambsganss 1998; Hattori et al. 1999; Mellier
1999).  
\item The third and oldest method uses optical spectroscopic
observations and determines the mass from the velocity distribution of
cluster galaxies with the assumption of virial equilibrium.
\end{itemize}

Many cluster masses have been determined in the three ways.
In some clusters there is agreement between the mass derived with the
different methods, in other clusters there are discrepancies typically
with the following relation 

\be
M_{X-ray} \approx M_{velocity~dispersion} \simlt
M_{weak~lensing}  \simlt M_{strong~lensing}
\label{eq:s1}
\ee
with differences sometimes as high as a factor of 2-3 between X-ray and
strong lensing mass. 
These discrepancies have studied extensively and
several possible explanations for them have been suggested:
\begin{itemize}
\item Observational uncertainties in the X-ray observations by ROSAT
and ASCA have been found by numerical simulations 
to be the main error source in the X-ray mass so far (Evrard et
al. 1996; Schindler 1996).
\item Non-equilibrium configurations, as are the case in merging
clusters, can lead to severe under- as well as overestimation of X-ray masses. 
\item Projection effects: lensing is sensitive to all the mass
along the line-of-sight, i.e. if there are mass concentrations in the
back- or foreground they are included in the lensing mass. The X-ray
mass determination is only sensitive to the potential well filled with
hot gas and measures therefore only the cluster mass.
\item Multi-phase medium (gas of different temperatures and densities
in pressure equilibrium), as is assumed to be present in cooling
flows, can lead to underestimation of the mass if only a single
temperature is used for the mass determination (Allen 1998).
\item Magnetic fields have also been suggested to be the reason for the
discrepancy. In the X-ray mass determination only thermal pressure is
taken into account. If there were in addition 
considerable magnetic pressure this would lead to an underestimation
of the mass. With magneto-hydrodynamic 
simulations, though, we have shown that the effect
of magnetic fields on the mass determination is essentially 
negligible (Dolag \&
Schindler 2000; Fig.~2).
\end{itemize}

\begin{figure}[thb]
 \begin{center}
  \mbox{\epsfig{file=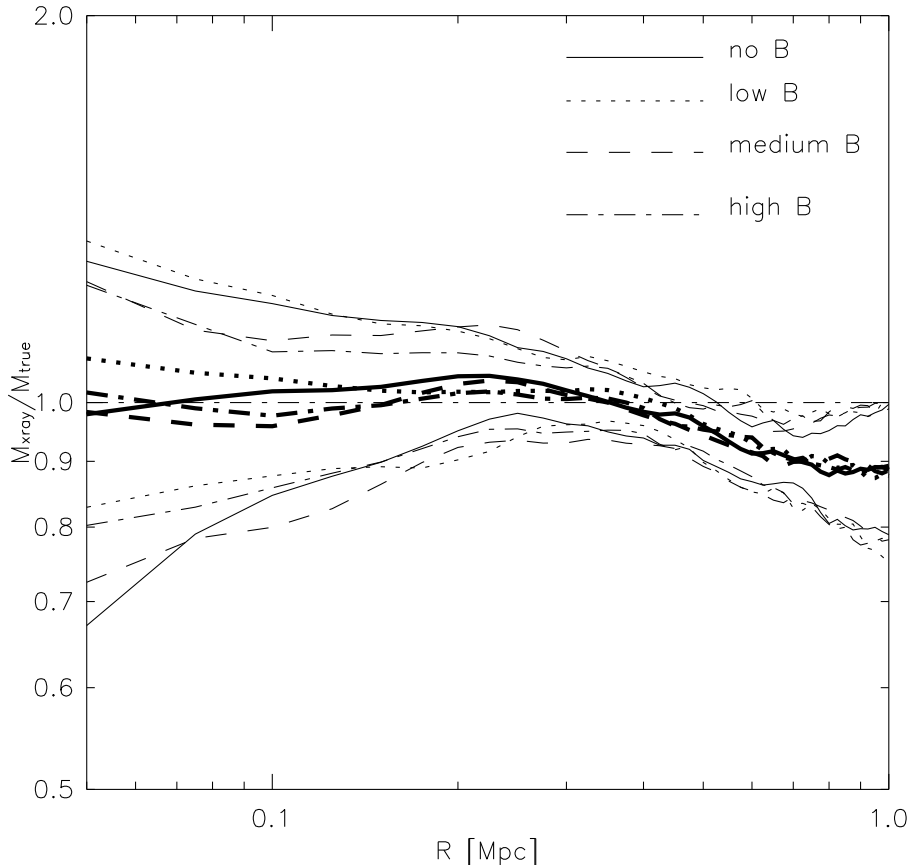,width=12cm,clip=}}
  \caption{\em Ratio of true and X-ray mass for 4 different
magnetic field strengths as seen in magneto-hydrodynamic simulations
(Dolag et al. 1999). 
The bold lines are averaged
profiles over 18 model configurations, while the thin lines show the
  corresponding standard deviation. 
Hardly any dependence of the mass estimate on the magnetic field
strength is visible (from Dolag \& Schindler 2000).
}
 \end{center}
\end{figure}

The ratio of visible to total mass
is a measure for the baryon fraction $\Omega_{baryon}$. 
Current measurements are in contradiction with primordial
nucleosynthesis for  $\Omega = 1$ (White et al. 1993), which is
therefore one of the important hints for a low $\Omega$ universe. 

The
relative distribution of gas and dark matter within a cluster gives also
useful information on physical processes occuring in clusters. The
fact that the gas is more extended than the dark matter -- an effect
which is stronger for the less massive clusters (see Fig.~3) -- can be
explained by additional (non-gravitational) heating processes, 
which are more efficient in
less massive clusters (Schindler 1999; Ponman et al. 1999). This gives
in principle hints on when and how the heat is produced and hence information
about cluster formation. But for stringent conclusions better data are
required (see Fig.~3).

\begin{figure}[thb]
 \begin{center}
  \mbox{\epsfig{file=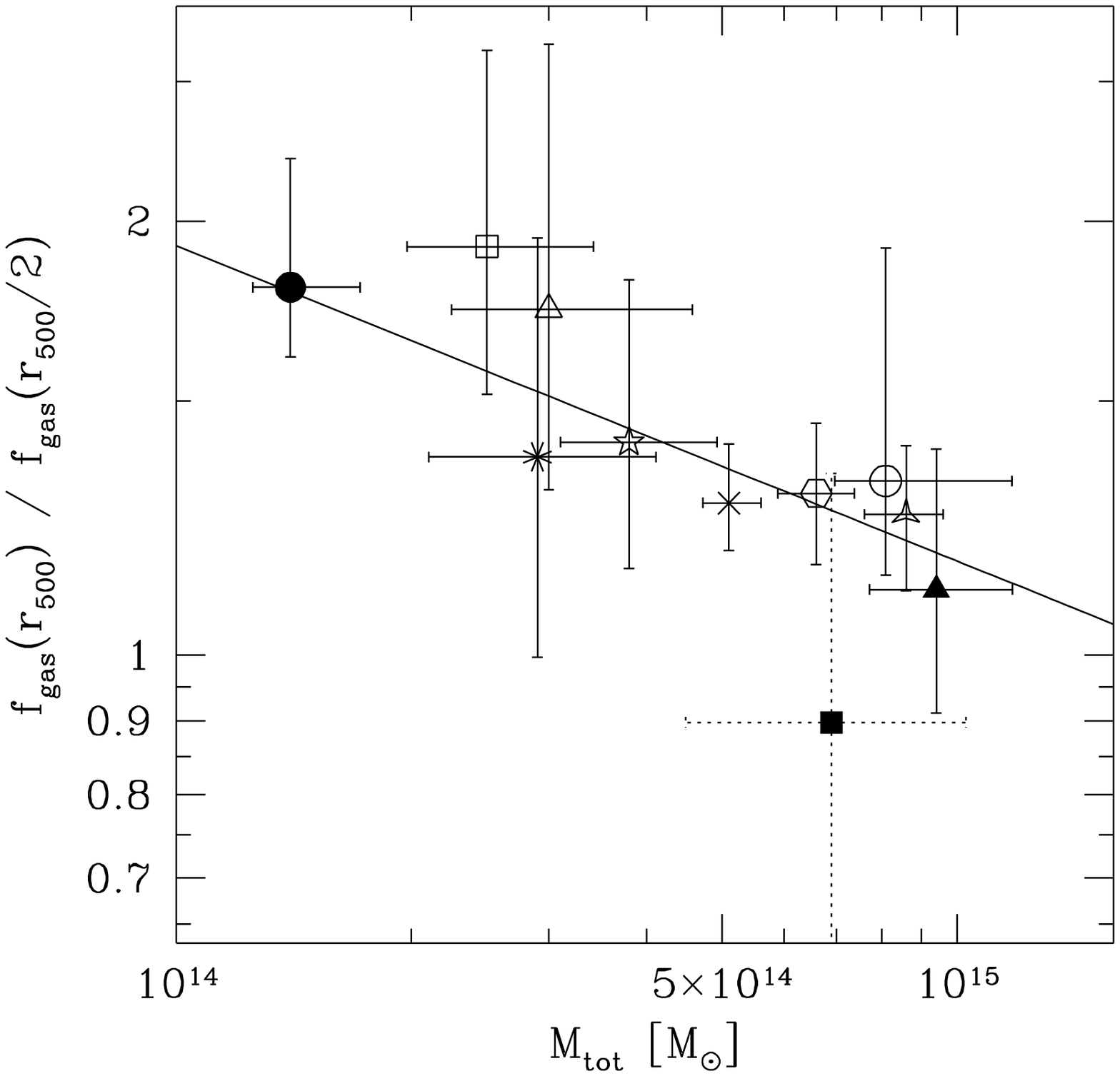,width=10cm}}
  \caption{\em {Gas extent relative to the dark
matter extent expressed as the ratio of gas mass fractions at large
radius and small radius (from Schindler 1999). 
The gas is relatively more extended in less
massive clusters, which is an indication of additional,
  non-gravitational 
heating processes.}}
 \end{center}
\end{figure}

The relatively large range found for the gas mass fraction 
does not only reflect the large observational uncertainties, but there
are really large variations in the gas mass fraction of individual clusters
(Ettori \& Fabian 1999; Schindler 1999; see Fig.~4). 
These large variations, which span 
an order of magnitude in extreme cases, have some implications for
cluster formation. If all the clusters had originally the same small
gas mass fraction and all the differences came later by different
amounts of gas released by the cluster galaxies, larger metallicities
in clusters with high gas mass fraction would be expected. But this is
not observed. Therefore the difference must be caused at least
partially by the primordial distribution of baryonic and non-baryonic
matter. 

The new X-ray and optical
facilities will certainly soon clarify many of these questions. 
Chandra and XMM with their improved spatial and spectral resolution
will provide very accurate measurements of cluster masses and hence
the amount and the distribution of dark matter.

\begin{figure}[thb]
 \begin{center}
  \mbox{\epsfig{file=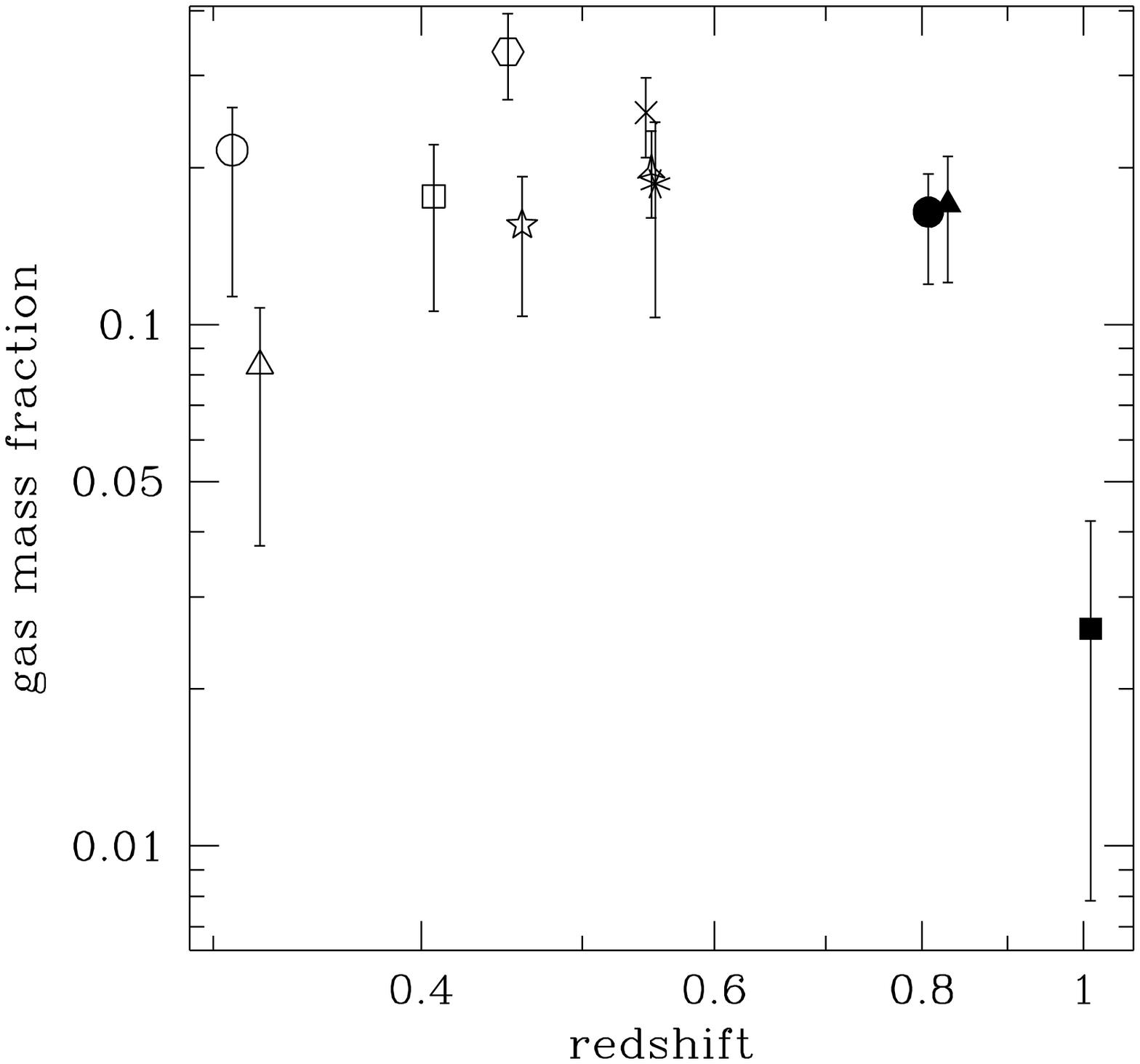,width=10cm}}
\vskip 0.7cm
  \caption{\em {
Gas mass fractions of various clusters versus redshift (from Schindler
  1999). There are
  large variations in the gas mass fraction from cluster to cluster. No
  significant trend of the fraction with redshift was found. 
}}
 \end{center}
\end{figure}


\section{Metal enrichment and the origin of the intra-cluster gas}

Spectroscopic X-ray observations of clusters show metal lines which
correspond to 
metallicities around 0.3 in solar units (Fukazawa et al. 1998).
This indicates that the gas cannot be purely of primordial origin.
Instead it must have been processed in the cluster galaxies, at least
partially, and then have been expelled from the galaxies' 
potential wells into the intra-cluster gas. 
The amount of this processed intra-cluster material
is not negligible, but equals or even exceeds (Mushotzky 1999) the
amount of metals in the galaxies themselves. There has been much
speculation about the processes 
responsible for the transport of gas from within the galaxies into the
intra-cluster gas and two mechanisms emerged as probable: galactic
winds (De Young 1978) and ram-pressure stripping (Gunn \& Gott 1972).
Light can be shed on which mechanisms operate by measuring
metallicity gradients and  metallicity evolution. 

Metallicity gradients can indicate what mechanism
prevails, because the different mechanisms expel the gas with different
radial distribution. 
Differences in the distribution of iron and
$\alpha$-elements give information on the type of supernovae the
metals origin from and hence on different enrichment time scales. 
So far, in several clusters
hints for radial metallicity gradients have been detected (e.g. 
Finoguenov et
al. 1999; De Grandi \& Molendi 1999a; Irwin \& Bregman 2000) from
observations with ASCA and BeppoSAX, but again for more detailed
results we have to wait for XMM and Chandra.

The time evolution of cluster metallicities is another obvious way to
investigate different processes, because they are expected to have
different time dependencies. Observations of clusters
out to $z\approx 0.5$ did not show any significant evolutionary
effects (Mushotzky \& Loewenstein
1997). Observations of high redshift clusters out to $z\approx 1$
(Schindler 1999) are consistent with this result, but are very
restricted due to large errors. In particular XMM with its high 
sensitivity will provide exact metallicity 
measurements for many distant clusters, which will
give important clues on cluster formation and galaxy formation.

\section{Dynamical state}

The dynamical state of clusters, whether
already relaxed or still consisting of merging subclusters, offers
another way to measure $\Omega$ as different scenarios predict
different evolution times (Richstone et al. 1992). 
While in some clusters the unrelaxed state is already obvious in the
X-ray images (e.g. in Abell 1437, see Fig.~5), for most clusters the
dynamical state is not so easy to determine.
It has been shown in 
numerical hydrodynamic simulations,
that the dynamical state can be characterised best by taking
into account not only the distribution of the X-ray emission but 
by comparing it also with the temperature distribution (Schindler \&
M\"uller 1993). Inhomogeneities in the temperature 
distribution arise in clusters during mergers: shock fronts emerge before
and after
the collision, which show up as temperature gradients. Also heated 
gas between two subclusters can be observed before the core passage.
From ASCA and BeppoSAX observations coarse temperature maps have been
produced 
(Markevitch et al. 1998; De Grandi \& Molendi 1999a,b; Irwin et al. 1999),
but the capabilities for spatially resolved spectroscopy 
of the new X-ray instruments on XMM and Chandra
promise well resolved temperature maps.

\begin{figure}[thb]
 \begin{center}
  \mbox{\epsfig{file=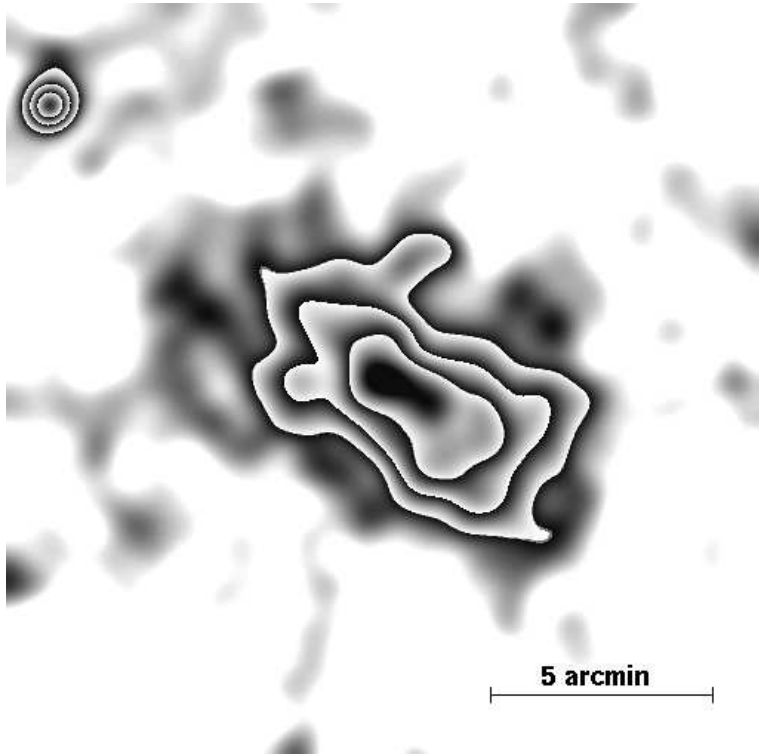,width=10cm,clip=}}
  \caption{\em {
ROSAT/HRI image of the cluster Abell 1437.
The emission of the cluster is strongly elongated in SW-NE
direction and the cluster centre in X-rays
does not coincide with the optical position -- both indications of a
  cluster in which subclusters are in the process of merging (from
  Schindler 2000).
The point source in the NE is
probably not connected with the cluster.
}}
 \end{center}
\end{figure}

Of particular interest for the determination of the dynamical state is
the comparison of X-ray observations with observations in other
wavelengths. The spatial distribution as well as the velocity
distribution of the galaxies obtained from optical observations
yield complementary
information about the morphology and kinematics. 
Radio emission can provide additional hints in two different
ways. (1) The relative
velocity between radio galaxies and the intra-cluster gas can be
inferred from the orientation and the shape of radio tails, so that the
kinematics perpendicular to the line of sight can be estimated. (2)
Diffuse, extended radio emission, which is not associated with a
particular galaxy -- radio halos and relics -- can be an
indication of a recent merging process (Giovannini et al. 1999; 
Feretti et al. 2000), because the shock waves emerging during the
collision of subclusters can reaccelerate particles to relativistic
energies, so that they emit synchrotron radiation.

\section{Conclusions}

Clusters of galaxies can be used for very different types of
cosmological tests. In particular observations with 
the new X-ray satellites launched
recently, XMM and Chandra, will provide a large step forward in
several respects. 
The unprecedentedly high spatial resolution of Chandra (1") provides
detailed X-ray images to investigate cluster morphology and derive
very accurate profiles.
The high sensitivity of XMM makes it the ideal instrument to find and study
distant galaxy clusters in order to investigate evolutionary effects.
Improved spectral resolution allows for much better temperature and
metallicity measurement.
And -- very importantly --
XMM and Chandra are capable of performing  
spatially resolved spectroscopy with high accuracy, which is necessary 
for temperature maps, mass determination and metallicity distributions. 
The combination of these data with observations in other wavelengths
will open new horizons in cosmology.

%
%

\section { References}

%
%

\reff Allen S., 1998, MNRAS 296, 392

\reff De Grandi S., Molendi S., 1999a, A\&A 351, L45

\reff De Grandi S., Molendi S., 1999b, ApJ 527, L25

\reff De Young D.S, 1978, ApJ 223, 47

\reff Dolag K., Bartelmann M., Lesch H., 1999, A\&A 348, 351

\reff Dolag K., Schindler S., 2000, A\&A in press

\reff Dressler A., 1980, ApJ 236, 351

\reff Ettori S., Fabian A.C., 1999, MNRAS 305, 834

\reff Evrard, A.E., Metzler, C.A., Navarro, J.N. 1996, ApJ, 469, 494

\reff Feretti L., Brunetti G., Giovannini G., Govoni F., Setti G.,
2000, astro-ph/0009346

\reff Finoguenov A., et al., 1999, astro-ph/9908150

\reff Fukazawa Y., Makishima K., Tamura T., 
    et al., 1998, PASJ 50, 187

\reff Giovannini G., Tordi M., Feretti L., 1999, New Astronomy 4, 141

\reff Gunn J.E., Gott J.R. III, 1972, ApJ 176, 1

\reff Hattori M., Kneib J.-P., Makino N., 1999, Prog. Theor. Phys.,
Supplement 133, 1 

\reff Irwin J.A., Bregman J.N., 2000, astro-ph/0009237

\reff Irwin J.A., Bregman J.N., Evrard A.E., 1999, ApJ 519, 518

\reff Markevitch M.L., Forman W.R., Sarazin C.L. and Vikhlinin
A., 1998, ApJ 503, 77

\reff Mellier Y., 1999, ARA\&A 37, 127

\reff Mushotzky R.F., 1999, astro-ph/9912547

\reff Mushotzky R.F., Loewenstein M., 1997, ApJ 481, L63

\reff Ponman T.J., Cannon D.B., Navarro J.F., 1999, Nature 397, 135

\reff Richstone D., Loeb A., Turner E.L., 1992, ApJ 393, 477

\reff Schindler S., 1996, A\&A, 305, 756

\reff Schindler S., 1999, A\&A 349, 435

\reff Schindler S., 2000, A\&AS  142, 433

\reff Schindler S., M\"uller E., 1993, A\&A 272, 137

\reff Schindler S., Binggeli B., B\"ohringer H., 1999, A\&A 343, 420

\reff Wambsganss J., 1998, Living Reviews in Relativity 1, 1998-12

\reff White S.D.M., Navarro J.F., Evrard A.E., Frenk C.S., 1993, Nat
      366, 429

\end{document}